# Multi-stage ytterbium fiber-amplifier seeded by a gain-switched laser diode

M. Ryser[1], S. Pilz[2], A. Burn[2], V. Romano[1,2]

[1]Institute of Applied Physics, University of Bern, Sidlerstrasse 5,
CH-3012 Bern, Switzerland
[2]Bern University of Applied Sciences, ALPS, Pestalozzistrasse 20,
CH-3400 Burgdorf, Switzerland

**Abstract**

We demonstrated all-fiber amplification of 11 ps pulses from a gain-switched laser diode at 1064 nm. The diode was driven at a repetition rate of 40 MHz and delivered 13 µW of fiber-coupled average output power. For the low output pulse energy of 325 fJ we have designed a multi-stage core pumped pre-amplifier in order to keep the contribution of undesired amplified spontaneous emission as low as possible. By using a novel time-domain approach for determining the power spectral density ratio (PSD) of signal to noise, we identified the optimal working point for our pre-amplifier. After the pre-amplifier we reduced the 40 MHz repetition rate to 1 MHz using a fiber coupled pulse-picker. The final amplification was done with a cladding pumped Yb-doped large mode area fiber and a subsequent Yb-doped rod-type fiber. With our setup we reached a total gain of 73 dB, resulting in pulse energies of >5.6 µJ and peak powers of >0.5 MW. The average PSD-ratio of signal to noise we determined to be 18/1 at the output of the final amplification stage.

**Introduction**

In laser materials micro-processing ultra-short laser pulses in the picosecond or femtosecond regime are used, when high demands on machining quality are posed. When processing metals, very good results concerning machining precision have been obtained with pulse lengths in the range of 10 ps [1], [2]. Under visual inspection, the lateral precision of the micro processed details does not improve significantly, when reducing the pulse length to 1 ps. However, studies show that the efficiency of material removal significantly increases when processing is done with pulses in the sub-picosecond range [3].

On the other side, when increasing the pulse length above 10 ps, a slight precision decrease due to an increase in heat affected zone is noticed [4], [5].

Considering ablation efficiency, workers in the field show that material ablation with pulse durations of 10 ps and less are more efficient in terms of volume ablation rate compared to pulse durations of >20 ps [1]. For shorter pulse durations, namely 0.1 ps-5 ps, it has been shown that ablation of iron (Fe) becomes even more efficient [3]. Processing of metals with nanosecond pulse durations allow to control the spatial ablation resolution at a length scale of tens of micrometers. With pico- and femtosecond-pulse durations material processing of structures on the nanometer length-scale can be achieved. This applies for the ablation resolution per pulse in depth. However, the achievable spot size of the beam is diffraction limited and thus in the order of the laser wavelength.

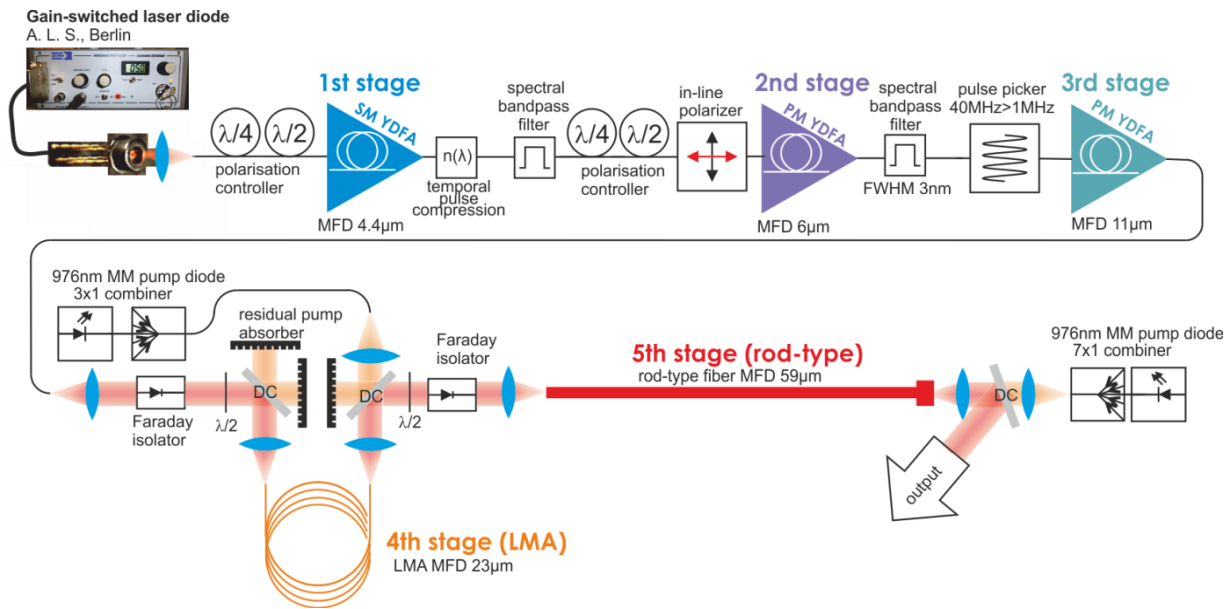

Figure 1: Experimental setup. The gain-switched laser diode seeded fiber amplifier consisted out of five amplification stages.

Considering these boundary conditions, a laser system with large material removal rate while maintaining high ablation resolution must deliver short pulse durations (10 ps and shorter), high peak power (≈MW) good beam quality ($M^2$≈1) and high pulse repetition frequency (≈MHz).

Here, we present the approach of combining a gain-switched laser diode [6] with an ytterbium doped fiber amplifier (YDFA) [7] for the generation of high energetic picosecond pulses.

Gain-switched laser diodes allow low-jitter electronic triggering of laser pulses. The drawback of using gain-switched laser diodes as seed for optical amplifiers is the weak output pulse energy in the order of few 100 fJ [6] and the therewith arising problem of maintaining a good signal to noise ratio during optical amplification. However, fiber amplifiers allow relatively low-noise amplification of weak signals and are thus a fist choice to amplify the weak pulses from gain-switched laser diodes [7].

Here, we also present preliminary results of signal to noise measurements that were done by using a novel time-domain method for determining the noise background of ultra-short pulse fiber amplifiers. The main contribution of noise in our fiber-optical amplifiers comes from amplified spontaneous emission (ASE), therefore we use henceforward the term ASE instead of noise. The method for determining the signal to ASE ratio is based on measuring the temporal convolution of the transmission window of an acousto-optic modulator (AOM) with the amplifier output that contains the pulse-train and the background ASE. By analyzing the data by the method of iterative re-convolution fitting, the signal and ASE content can be discriminated from each other. The method will be presented in detail elsewhere; here we present first results of this method.

**Experimental Setup**

The experimental setup of our fiber-laser system is sketched in Fig. 1. The gain-switched laser-diode (Advanced Laser Diode Systems, Berlin) delivered 13 µW of average fiber coupled optical output power. The optical output of the gain-switched laser-diode were un-polarized picosecond pulses at a wavelength of 1063.3 nm and a bandwidth of 0.6 nm (full width, cp. Fig. 2), with 32 ps temporal width (FWHM, cp. Fig. 3) and at a repetition rate of 40 MHz. With a dispersion compensating element in the first amplification stage, the pulses were temporally compressed. After the first stage the signal passed an inline-polarizer and the subsequent amplification was based on polarization maintaining optical components and fibers. After the second amplification stage, the repetition rate was reduced from 40 MHz to 1 MHz for the final amplification stages. The core diameters of the optical fibers were gradually increased for the consecutive amplification stages up to 70 µm. This prevented undesired non-linear effects or even damage of the fibers due to the high pulse peak powers. The two amplification stages were done by using a large mode area (LMA) fiber with 30 µm and a rod-type fiber of 70 µm core diameter, both optically pumped at a wavelength of 976 nm in counter-pump configuration.

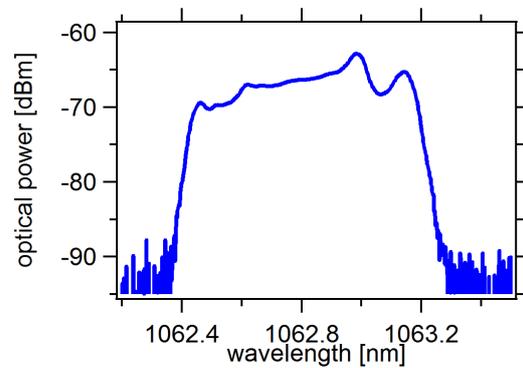

Figure 2: Spectral characteristics of the gain-switched laser diode. The maximum spectral emission of the gain-switched laser diode was at 1063 nm and the full bandwidth was 0.6 nm

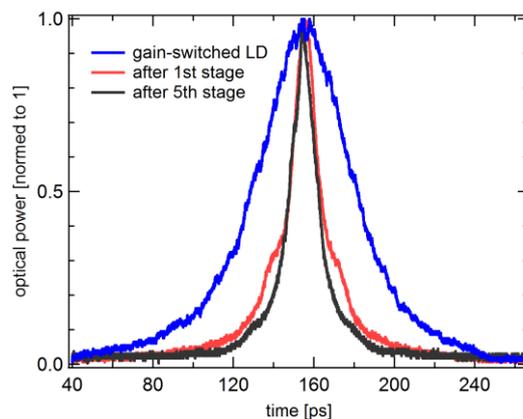

Figure 3: Auto-correlator traces of the picosecond-pulses at three positions. blue: seed pulse generated by the gain-switched laser diode, red: after passing the dispersion compensating element and first amplification stage, black: pulse shape after amplification through all amplification stages.

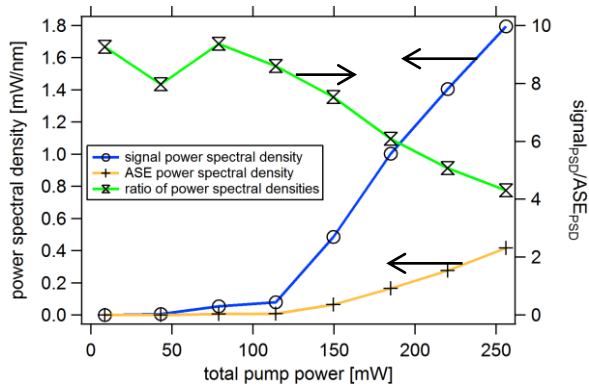

Figure 4: Signal/ASE performance of the first amplification stage.

## Results

Each YDFA-stage was optimized for high gain and low ASE contribution.

Depending on the chosen polarization plane, multiple pulses separated by few picoseconds or single pulses were observed after the in-line polarizer. Thus, the polarization controllers shown in Fig. 1 were adjusted to get single pulses after linear polarization. The temporal compression by the dispersion compensating element (Fig. 1) was about a factor of 3, resulting in a temporally compressed laser pulse of about 11 ps FWHM. The auto-correlator measurements in Fig. 3 show the pulse shape before and after temporal compression.

Due to the low seed power provided by the gain-switched laser diode, it was very important to find an optimal working point of the fiber-pre-amplifier, where high gain at low ASE contribution could be achieved. By using our new time-domain approach that allowed us to determine the signal and ASE content, we identified the optimal working point of the first amplification stage. The average power spectral densities (PSD) of the signal and the amplified spontaneous emission over the 0.6 nm bandwidth of the signal in dependence of pump power at 976 nm are shown in Fig. 4. The ratio of the two PSDs indicates that at the optimal working point the ratio of signal/ASE is about 9/1. Note, that the temporally averages signal/ASE ratio is further improved to 18/1 by temporal filtering in the AOM between the 2nd and 3rd stage (cp. Fig. 1).

The second YDFA-stage was gain-limited by the occurrence of temporal pulse-width broadening. After the second YDFA-stage the repetition rate was reduced from 40 MHz to 1 MHz by using a pulse-picker. The achievable gain in the third YDFA-stage was also gain-limited by the occurrence of temporal pulse broadening. For both stages we observed a rising peak at approximately 1217 nm. This is in the wavelength region, where a Raman-peak is expected to arise for such fibers.

The fourth stage was gain-limited by the occurrence of spontaneous lasing. We assume these came from weak back-reflections of the angle-cleaved LMA fiber-ends. In the fifth (rod-type) YDFA-stage we achieved an average signal output power of 5.6 W at a repetition rate of 1 MHz. This corresponds to a pulse energy of 5.6 µJ and a peak power of >0.5 MW. The gain in the fifth YDFA-stage was limited by available pump power.

The overall performance of our system is shown in Fig. 5 and Fig. 6. Fig. 5 illustrates the spectral broadening of the signal during the amplification process. The traces in Fig. 6 depict the amplification characteristics and the achieved peak power for each amplification stage. The total gain of all amplification stages was 73 dB. The pulse shapes did not change significantly throughout the amplification process as indicated by the auto-correlation traces shown in the Fig. 3.

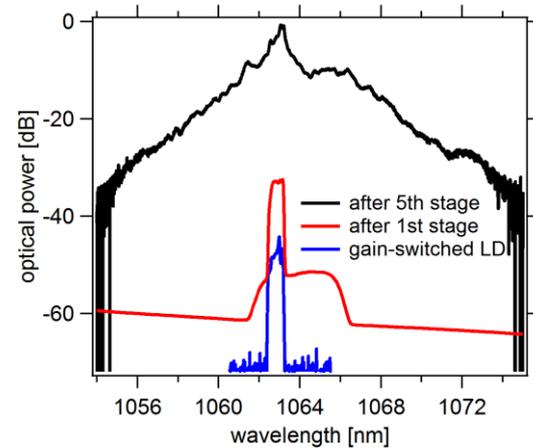

Figure 5: Spectral characteristics of signal being amplified. Blue: seed pulse generated by the gain-switched laser diode, red: after passing the dispersion compensating element, first amplification stage and band-pass filter, black: spectrum of the signal after amplification through all amplification stages.

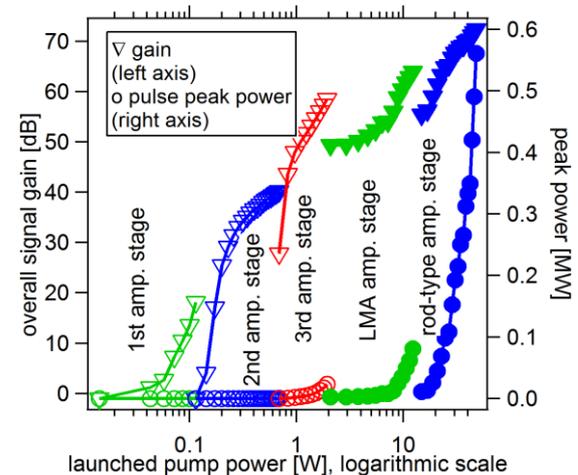

Figure 6: Amplification characteristics of all five stages. The graph shows the output measured after each YDFA-stage. Note that the gain-curves respectively the pulse peak power curves of a later YDFA-stage starts below the preceding output because in between the stages were placed optical elements (cp. Fig. 1) reducing the transmitted power.

## Summary

The preamplifiers were optimized towards low ASE contribution. By using our new approach of determining the signal to ASE ratio in time-domain, we could optimize the pre-amplifiers for low ASE contribution (Fig. 4). The average PSD-ratio of signal to ASE was determined to be 18/1 at the output of the final amplification stage.

The output of a gain-switched laser diode was compressed approximately by a factor of three by using a dispersive optical element (cp. Fig. 1) to a pulse-duration of 11 ps (Fig. 3).

We have demonstrated the amplification of the weak pulses (0.26 pJ) by 73 dB to an output pulse energy of >5.6 μJ (Fig. 6). After optimizing the multi-stage fiber amplifier, no significant temporal pulse broadening was observed during the amplification process (Fig. 3).

We estimated the final peak power to be >0.5 MW. To the best of our knowledge, we achieved the shortest pulse duration and highest peak power with ytterbium fiber based direct amplification of laser pulses generated by a gain-switched laser diode that is driven at MHz repetition rates.

## References

[1] B. Jaeggi, B. Neuenschwander, M. Schmid, M. Muralt, J. Zuercher, and U. Hunziker, "Influence of the Pulse Duration in the ps-Regime on the Ablation Efficiency of Metals," Physics Procedia, vol. 12, no. 2010, pp. 164–171, 2011.

[2] B. Neuenschwander, B. Jäggi, M. Schmid, U. Hunziker, B. Luescher, and C. Nocera, "Processing of industrially relevant non metals with laser pulses in the range between 10ps and 50ps," in ICALEO 2011, 2011, p. Paper M103.

[3] N. N. Nedialkov, S. E. Imamova, and P. a Atanasov, "Ablation of metals by ultrashort laser pulses," Journal of Physics D: Applied Physics, vol. 37, no. 4, pp. 638–643, Feb. 2004.

[4] Y. Hirayama and M. Obara, "Heat-affected zone and ablation rate of copper ablated with femtosecond laser," Journal of Applied Physics, vol. 97, no. 6, p. 064903, 2005.

[5] X. Zhu, "A new method for determining critical pulse width in laser material processing," Applied surface science, vol. 167, no. 3–4, pp. 230–242, 2000.

[6] P. P. Vasil'ev, "Ultrashort pulse generation in diode lasers," Optical and Quantum Electronics, vol. 24, no. 8, pp. 801–824, Aug. 1992.

[7] R. Paschotta, J. Nilsson, and A. Tropper, "Ytterbium-doped fiber amplifiers," , IEEE Journal of, vol. 33, no. 7, pp. 1049–1056, 1997.